\documentstyle[12pt,colordvi,psfig]{article}

\begin{document}
\renewcommand{\thefootnote}{\fnsymbol{footnote}}
\begin{center}
{\LARGE {\bf Langevin dynamics of $J/\psi$ in a parton plasma}}\vspace{.5in}\\
{\large B. K. Patra$^a$
\footnote{Present address: Saha Insitute
of Nuclear Physics, 1/AF Bidhan Nagar, Kolkata 700 064}
 and V. J. Menon$^b$} \vspace{.5in}\\
{\it $^a$ Variable Energy Cyclotron Centre, 1/AF Bidhan Nagar, Calcutta 700 064,
India\\
$^b$ Department of Physics, Banaras Hindu University, Varanasi 221 005, India\\
}
\vspace{.7in}
\underline{\large Abstract} 
\end{center}
We consider the Brownian motion of a $c \bar c$ pair produced in the very
early satge of a quark-gluon plasma. The one-dimensional Langevin equation
is solved formally to get purely mechanical properties at small and large
times. Stochastically-averaged variances are examined to extract the time
scales associated with swelling and ionization of the bound state. Simple
numerical estimates of the time scales are compared with other mechanisms
of $J/\psi$ suppression.\\

PACS number(s): 12.38.Mh, 05.10.Gg, 52.65.Ff, 05.20.Dd
\pagebreak

\section {\bf {Introduction}}
\setcounter{equation}{0}
\renewcommand{\theequation}{2.\arabic{equation}}

Charmonium i.e., $J/\psi$ suppression~\cite{matsui} continues to be one of the
most hotly debated signatures of the production of a quark-gluon plasma
in ultrarelativistic heavy ion collisions. Both the initial formation
and subsequent survival probabilities of the $J/\psi$ are affected by
several factors {\em viz.} scattering with hard partons~\cite{xu}, Debye colour
screening~\cite{matsui}, drag and diffusion arising from Brownian 
motion~\cite{sve,munshi},
evolution of the plasma  via hydrodynamic flow~\cite{pal}, etc. Although the
theoretical time scales for gluonic dissociation {\em vs} colour
screened break-up are well known~\cite{binoy} the time scales for the
swelling/ionization of $J/\psi$ caused by Brownian movement are not yet 
understood
satisfactorily, and the aim of the present paper is to focus attention
on this aspect.

The classical $3$-dimensional Fokker-Planck equation for the distribution
function of a {\it single} charmed quark propagating in a plasma was
first studied by Svetitsky~\cite{sve}. Assuming soft scattering with the partons
he found the associated Boltzmann transport coefficients for drag
and diffusion to be rather large but he ignored the force
which binds the $c\bar  c$ pair. Plotnik and Svetitsky~\cite{daphne} extended this
philosophy to the case of {\it two-particle} Fokker-Planck dynamics in the
presence of a colour singlet/octet potential between the pair. However,
since no attempt was made by them to actually solve the resulting $13$
variable partial differential equation, hence no simple estimate
was given for the time scales of the $c \bar c$ pair.

In the present work  we adopt a different approach based on analyzing the
one-dimensional Langevin equation for the stochastic trajectory of a
$c \bar c$ pair initially bound by a screened Coulomb field. It is
known since long ago~\cite{chandra} that the Langevin theory provides a valid
description of classical Brownian movement of a test particle acted
upon simultaneously by a driving interaction, frictional force, thermal
agitation, and random noise. In Sec.2 below we write formal
solutions to the underlying equations of motion and obtain
compact expressions for the purely mechanical observables
at small and large times. Stochastic averaging of the observables 
is done in Sec.3 so as to deduce statistical properties
({\em viz.} means, variances, time scales, etc.) of the
system. Sec.4 gives simple, order-of-magnitude estimates
of the relevant time scales and discusses the result {\em vis-a-vis}
other mechanisms of $J/\psi$ suppression. Finally, Sec.5 examines
critically the validity of our main assumptions and also
mentions 
several additional complications which would have to be incorporated
in future applications of the theory.

\section {\bf {Purely Mechanical Observables}}
\setcounter{equation}{0}
\renewcommand{\theequation}{2.\arabic{equation}}
{\bf 2A. Assumptions \& notations}

For a nonrelativistic $c \bar c$ pair propagating in a spatially homogeneous
plasma the net external colour force due to the background is
zero although the internal screened Coulomb potential (assumed to be
colour singlet) survives. Working in the barycentric frame and
adopting a one-dimensional view we can describe the motion of an
effective single particle by defining
\begin{eqnarray}
&&t={\mbox{\rm{time}}}, \quad m = {\mbox{\rm{reduced mass}}},\quad x ={\mbox 
{\rm{position}}}, \quad u=\dot{x}={\mbox{\rm{velocity}}},\nonumber\\
&&V=V(x)={\mbox{\rm{binding potential}}}, \quad f =f(x)=-\partial V /\partial x = 
{\mbox{\rm{binding force}}}, \nonumber\\
&&\gamma = {\mbox {\rm{coefficient of friction (per unit mass) assumed 
constant}}}, \nonumber\\
&&T = {\mbox{\rm{ambient temperature of medium in energy units}}},\nonumber\\
&&h = h(t)= {\mbox {\rm{random force taken as a white Gaussian noise}}}, 
\nonumber\\
&&C = 2 m \gamma T = {\mbox {\rm{strength of the noise autocorrelation 
function}}},\nonumber\\
&&D = C/2 m^2 \gamma^2 = T/m \gamma = {\mbox{\rm{Einstein diffusion coefficient}}}
\end{eqnarray}

The space derivative $f^\prime$ and total time derivative $\dot f$ of the
force are written as
\begin{eqnarray}
f^\prime = \partial f /\partial x, \quad \dot {f}= df/dt = u f^\prime
\end{eqnarray}
Since the pair potential $V$ depends upon $|x|$ the symmetry
relations
\begin{eqnarray}
V(-x) = V(x), \quad f(-x) = -f(x)
\end{eqnarray}
hold as shown schematically in Fig.1

\begin{figure}
\psfig{file=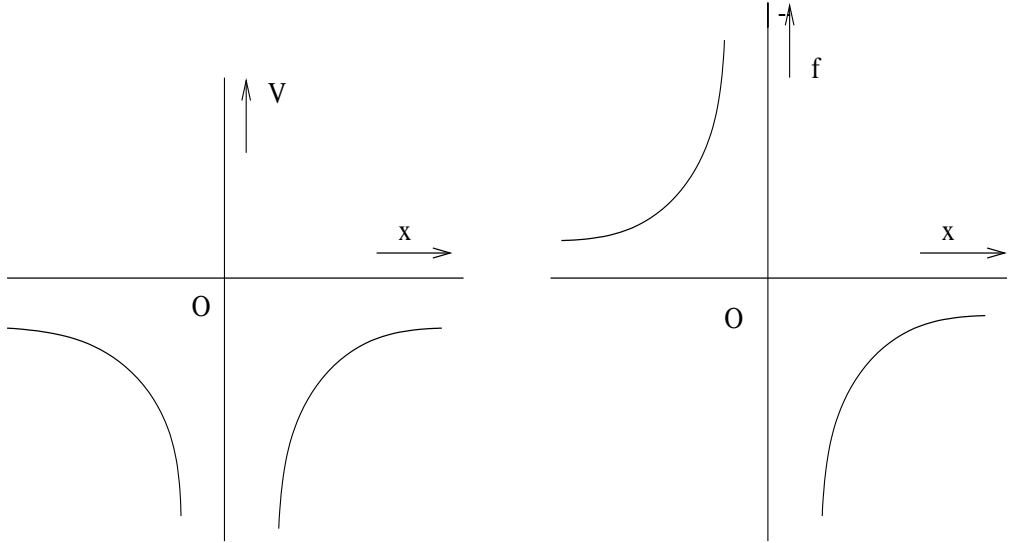,angle=0,height=8cm,width=14cm}
\vskip 0.1in
\caption{ Schematic shapes of $c \bar c$ binding
potential $V$ and force $f$.
}
\end{figure}

Before proceeding further a few important comments are in order. 
One-dimensional stochastic models~\cite{chandra,kamper} have been found very useful
in  the past because they are mathematically simple and can also
simulate purely radial motion in three dimensions. The choice $C=2m\gamma T$
guarantees that the test particle's distribution at asymptotic time
would become Maxwell-Boltzmann at background temperature $T$ in accordance
with the fluctuation-dissipation theorem~\cite{kubo}. Due to the same reason
the single diffusion  parameter $D = T/m\gamma$ gets fixed in terms of the
temperature and the damping coefficient. All these remarks are, however, 
subject to alterations as will be pointed out later in Sec. 5.\vspace{.2in}\\

{\bf 2B. Langevin equation \& the velocity}

The basic ordinary differential equation to be considered is
\begin{eqnarray}
\dot {u} + \gamma u = (f +h)/m, \quad \dot {x} = u
\end{eqnarray}
subject to the initial conditions
\begin{eqnarray}
u(t=0) =u_0, \quad x(t=0) =x_0
\end{eqnarray}

For a free particle and a harmonic oscillator, Eq.(2.4) was solved explicitly
by Chandrasekhar~\cite{chandra} but the present case is more difficult because $f$
is a nonlinear function of $x$. Using the integrating factor $e^{\gamma t}$
we obtain a formal solution for the velocity as
\begin{eqnarray}
u = u^{\mbox{\rm{fr}}} + v + w
\end{eqnarray}
Here the contributions arising from free Rayleigh motion, the driving force,
and the random noise are respectively given by
\begin{eqnarray}
u^{\mbox{\rm{fr}}}&=&u_0 e^{-\gamma t}, \quad v = \frac{e^{-\gamma t}}{m} \int_{0}^{t}
dt_1 e^{\gamma t_1} f_1 \nonumber\\
w&=&\frac{e^{-\gamma t}}{m} \int_{0}^{t} d t_1 e^{\gamma t_1} h_1
\end{eqnarray}
with $t_1$ being an integration time and $f_1=f\mid_{t_1}, \quad h_1 = h\mid_{t_1}$.

At small times the driving force can be approximated by a first-order Taylor
expansion
\begin{eqnarray}
f_1 \approx f_0 + u_0 f_0^\prime t_1 + \cdot \cdot \cdot, \quad \quad \gamma t_1 \ll 1 
\end{eqnarray}
where the suffix $0$ refers to the instant when the $J/\psi$ was
produced and the dots represent nonleading terms. Substituting into
Eqs.(2.6), (2.7) we get the initial behaviour of the velocity and its 
square as
\begin{eqnarray}
u &\approx & u_0 + J_0 t + \cdots +w \quad, \quad \quad t \ll \gamma^{-1} \\
u^2 & \approx & \{ u_0^2 + 2 u_0 J_0 t + \cdot \cdot \cdot \} + \nonumber\\
 &+& 2 \{ u_0 +J_0 t + \cdot \cdot \cdot \} w + w^2
\end{eqnarray}
with 
\begin{eqnarray}
J_0 = f_0/m - \gamma u_0
\end{eqnarray}
Since the piece $w$ containing noise may fluctuate rapidly with time
its Taylor expansion is not attempted. Next, at large times the
piece $v$ in Eq.(2.7) can be integrated by parts once to yield the Rayleigh 
estimate
\begin{eqnarray}
v \approx f/m\gamma + \cdot \cdot \cdot ,\quad \quad  \gamma t \gg 1
\end{eqnarray}
But the asymptotic value of $f$ is essentially zero both for a
bound system (where the particle tends towards a point of stable
equilibrium) as well as unbound one (where the particle tends to
fly away). Therefore, we arrive at the leading asymptotic behaviours
\begin{eqnarray}
u \approx w + \cdot \cdot \cdot, \quad u^2 \approx w^2 + \cdot \cdot \cdot, \quad
\quad t \gg \gamma^{-1}
\end{eqnarray}

{\bf 2C. Analysis of Langevin trajectory}

Integrating Eqs.(2.6), (2.7) with respect to $t$ we obtain the position, 
i.e., the relative separation between the $c \bar c$ pair
\begin{eqnarray}
x = x^{\mbox{\rm{fr}}} + y + z
\end{eqnarray}
Here the contributions arising from free Rayleigh motion, the driving
force, and the random noise are respectively read-off from
\begin{eqnarray}
x^{\mbox{\rm{fr}}} & =&  x_0 + \frac{u_0}{\gamma} \left( 1 - e^{-\gamma t} \right)
,\nonumber\\
y& =& \frac{1}{m \gamma} \int_{0}^{t} d t_1 K_{tt_1} f_1, \nonumber\\
z& =& \frac{1}{m \gamma} \int_{0}^{t} d t_1 K_{tt_1} h_1 
\end{eqnarray}
with $K_{tt_1}$ being a useful kernel defined by
\begin{eqnarray}
K_{tt_1} = 1 - e^{-\gamma (t -t_1)}
\end{eqnarray}
At small times the Taylor expansion (2.8) of the force can be used to
deduce the following {\it initial behaviour} of the trajectory and its square
\begin{eqnarray}
x & \approx & x_0 + u_0 t + \frac{J_0}{2} t^2 + \frac{K_0}{6}t^3+\cdot 
\cdot \cdot +z, \quad \quad t \ll \gamma^{-1} \\
x^2 & \approx & \{ x_0^2 + 2 x_0 u_0 t + ( u_0^2 +x_0 J_0) t^2 + 
(u_0J_0 + \frac{x_0K_0}{3} ) t^3 + \cdots \}  \nonumber\\
&&+ 2 \{ x_0 + u_0 t + \frac{J_0}{2} t ^2 + \frac{K_0}{6}t^3+\cdots \}
z + z^2 
\end{eqnarray}
where 
\begin{eqnarray}
K_0 = \frac{f_0^\prime u_0}{m} - \frac{\gamma f_0}{m} + \gamma^2 u_0
\end{eqnarray}
At large times the sum 
$x^{\mbox{\rm{fr}}}+y$ tends to the quantity
\begin{eqnarray}
X_\infty = x_0 + \frac{u_0}{\gamma} + \frac{1}{m \gamma} \int_{0}^{\infty}
dt_1 f_1, \quad \quad \gamma t \gg 1
\end{eqnarray}
whose value, however, is not known apriori. For a bound system $X_\infty$
may coincide with a point of stable equilibrium in the field $f$. However,
if the particle becomes unbound then $X_\infty$ may look like
$ x_0 +u_\infty^{\mbox{\rm{fr}}}/\gamma$ with $u_\infty^{\mbox{\rm{fr}}}$ being the final velocity
of free Rayleigh motion. Thus we obtain at the {\it asymptotic behaviour}
\begin{eqnarray}
x &\approx& X_\infty + z ,\nonumber\\
x^2 &=& X_\infty^2 + 2 X_\infty z + z^2 ,\quad \quad t \gg \gamma^{-1}
\end{eqnarray}
where no asymptotic expansion is attempted for the fluctuating term
$z$.\vspace{.2in}

{\bf 2 D. Treatment of Langevin energy}

Finally we turn to the mechanical energy $E=mu^2/2 + V$. Remembering the
dash-dot notation specified by Eq.(2.2) the rate of change of $E$
becomes
\begin{eqnarray}
\dot {E} &=& m u \dot{u} + V^\prime u \nonumber\\
&=& \left( h - m \gamma u \right) u
\end{eqnarray}
whose formal solution is
\begin{eqnarray}
E = E_0 + \int_{0}^{t} dt_1 \left( h_1 u_1 - m \gamma u_1^2 \right)
\end{eqnarray}
At small times the initial behaviour (2.9 - 2.10) of the velocity can be 
inserted into Eq.(2.23) to yield
\begin{eqnarray}
E \approx  E_0 + \left[ (u_0 +w)h - m \gamma {(u_0+w)}^2 \right] t 
+ \cdot \cdot \cdot \quad , \quad \quad t \ll \gamma^{-1}
\end{eqnarray}
At large times if the system remains bound then $E$ tends to a value below
the ionization thresholds. However, if the system does ionize then
the velocity $u$ tends to  $w$ while the short-range potential
$V$ approaches zero. Hence, for disintegrated pair
\begin{eqnarray}
E \approx m w^2/2 + \cdot \cdot \cdot \quad, \quad \quad t \gg \gamma^{-1}
\end{eqnarray}
Eqs.(2.7 - 2.24) describe the main, purely mechanical, properties of
interest to us; many of these expression may be regarded as new for general
shape of the driving force $f$.

%%%%%%%%%%%%%%%%%%%%%%%%%%%%%%%%%%%%%%%%%%%%%%%%%%%%%%%%%%%%%%%%%%%
\section {\bf {Stochastic Averaging}}
\setcounter{equation}{0}
\renewcommand{\theequation}{3.\arabic{equation}}
%%%%%%%%%%%%%%%%%%%%%%%%%%%%%%%%%%%%%%%%%%%%%%%%%%%%%%%%%%%%%%%%%%%

{\bf 3 A. Statistical input}

In analogy with Chadrasekhar's work~\cite{chandra} stochastic means will now
be taken with respect to the initial velocity $u_0$, initial position
$x_0$, and the Gaussian distributional of the noise $h$. 
For a genuine bound state obeying the virial theorem the
input expectation values read
\begin{eqnarray}
&&\langle u_0 \rangle =0,\quad \langle u_0^2 \rangle =\Delta_{u0}^2,
\quad \langle x_0 \rangle =0 \nonumber\\
&&\langle x_0^2 \rangle =\Delta_{x0}^2,\quad \langle x_0 u_0 \rangle =0, 
\langle x_0 f_0 \rangle /m = - \langle u_0^2 \rangle \nonumber\\
&& \langle h_1 h_2 \rangle = C \delta (t_1 -t_2)
\end{eqnarray}
where $\Delta_{u0}$ and $\Delta_{x0}$ are the velocity spread and
position spread, respectively, in the barycentric frame of the
$J/\psi$ produced at $t=0$. At general $t$ the fluctuating
velocity piece $w$ and fluctuating position piece $z$ have
the properties
\begin{eqnarray}
&&\langle w \rangle = 0, \quad \langle z \rangle = 0, \quad \langle w^2 \rangle = 
\frac{C}{2 \gamma m^2} \left( 1 - e^{-2\gamma t} \right), \nonumber\\ 
&& \langle z^2 \rangle = \frac{C}{m^2 \gamma^2} \left( t -
\frac{2}{\gamma} ( 1 - e^{-\gamma t} ) + \frac{1}{2\gamma}
( 1 - e^{-2 \gamma t} ) \right)
\end{eqnarray}

The informations (3.1) and (3.2) will be utilized below. \vspace{.2in}\\

{\bf 3 B. Initial behaviour of averaged observables}

Let us return to Eqs.(2.9), (2.10), (2.17), (2.18), (2.24) describing the 
velocity, position, and energy at small
times. Clearly $\langle J_0 \rangle =0 $ and $\langle K_0 \rangle = 0$ 
since $f_0$ is an odd function of $x_0$ in view of the assumption (2.3). 
Therefore, using the inputs (3.1 - 3.2) we deduce
\begin{eqnarray}
&&\langle u \rangle = 0, \quad \Delta_u^2=\langle u^2 \rangle  
\approx \Delta_{u0}^2 + g_u t + \cdot \cdot \cdot \nonumber\\
&&\langle x \rangle = 0, \quad  \Delta_x^2=\langle x^2 \rangle 
\approx \Delta_{x0}^2 + I_x t^3 + \cdot \cdot \cdot \nonumber\\
&& \langle E \rangle \approx \langle E_0 \rangle + g_{{}_E} t + \cdot 
\cdot \cdot \quad,\quad \quad \quad t  \ll \gamma^{-1}
\end{eqnarray}
where $\Delta_u$ is the velocity spread, $\Delta_x$ is the position
spread, the dots stand for nonleading terms, and the extra 
coefficients written are
\begin{eqnarray}
g_u &=& \left( C - 2 \gamma m^2 \Delta_{u0}^2 \right)/m^2 = 2 \gamma \left( 
T/m - \Delta_{u0}^2 \right) \nonumber\\
I_x &=& \left( C + 2 \gamma m \langle x_0 f_0 \rangle \right)/3m^2 
=g_u/3 \nonumber\\
\langle E_0 \rangle &=& m \Delta_{u0}^2/2 + \langle V_0 \rangle, \quad g_{{}_E} = mg_u/2
\end{eqnarray}
Eqs.(3.3 - 3.4) have four important physical consequences :

(i) The linear time-dependence of the velocity variance $\Delta_u^2$
is controlled by the coefficient $g_u$. Clearly $\Delta_u^2$ increases
with $t$ if $g_u > 0$ i.e. if $T > m \Delta_{u0}^2$ which is a
{\it nonequilibrium} situation. However, $\Delta_u^2$ remains constant in time
if $g_u =0$ i.e. if $T= m \Delta_{u0}^2$ which is an equilibrium situation.

(ii) The cubic time-dependence of the position variance $\Delta_x^2$
is governed by the parameter $I_x$. Evidently $\Delta_x^2$ increases with $t$
if $I_x > 0$ and the increment $I_x t^3$ becomes comparable
to the initial value $\Delta_{x0}^2$ at a time $\tau_x$ satisfying
$I_x \tau_x^3 = \Delta_{x0}^2$. In other words, Brownian movement can
cause the bound state to {\it swell} substantially within a time scale
\begin{eqnarray}
\tau_x = {\left( \Delta_{x0}^2/ I_x \right)}^{1/3} \quad < \gamma^{-1} \quad 
\quad {\mbox{\rm{if}}} \quad I_x > 0
\end{eqnarray}
(iii) The linear time-dependence of average energy $\langle E \rangle$
is controlled by the constant $g_{{}_E} = m g_u/2$. Obviously, $\langle E
\rangle$ increases with $t$ if $g_{{}_E} > 0$  i.e. if $T > m \Delta_{u0}^2$
and it becomes zero at a time $\tau_{{}_E}$ satisfying $g_{{}_E}
 \tau_{{}_E} = -
\langle E_0 \rangle$. In other words, Langevin dynamics can cause
$c \bar c$ to {\it ionize} after a time span
\begin{eqnarray}
\tau_{{}_E} = - \langle E_0 \rangle /g_{{}_E} \quad \quad {\mbox{\rm{if}}} 
\quad g_{{}_E} > 0
\end{eqnarray}
(iv) At this stage a couple of remarks must be added on the situation
where inequalities on the coefficients get reversed, i.e.
\begin{eqnarray}
I_x <0  \quad {\mbox {\rm{or}}} \quad g_{{}_E} < 0
\end{eqnarray}
The possibility $I_x < 0$ physically implies that the contraction
caused by the force term $\langle x_0 f_0 \rangle$ dominates over the 
expansion caused
by noise $C$ in Eq.(3.4) so that swelling of the bound state is 
{\it ruled out}. Next, the possibility $g_{{}_E} <0$ in Eq.(3.3) implies that the mean
energy $\langle E \rangle $ becomes deeper than $\langle E_0 \rangle $ which
again forbids the break-up of the classical bound state. Of course, the 
results contained in Eqs.(3.3 - 3.7) are new and original for a general shape of
the binding force $f(x)$. \vspace{.2in}\\

{\bf 3C. Asymptotic behaviour of averaged observables}

Let us assume that $J/\psi$ has disintegrated under the above-mentioned
conditions and examine Eqs.(2.13), (2.21), (2.25) describing mechanical 
properties at large times. Employing the statistical inputs (3.1 - 3.2) one finds
\begin{eqnarray}
\Delta_x^2 & \sim & C t/m^2 \gamma^2 + \cdot \cdot \cdot \quad \sim 2 D t + \cdot \cdot 
\cdot \nonumber\\
\Delta_u^2 &\sim & C/2 \gamma m^2 + \cdot \cdot \cdot \quad \sim T/m + \cdot \cdot 
\cdot \nonumber\\
\langle E \rangle &\sim & T/2 + \cdot \cdot \cdot \quad, \quad \quad t \gg \gamma^{-1}
\end{eqnarray}
where the contribution to $\Delta_x^2$ arising from the variate $X_\infty$
has been omitted. Of course, Eq.(3.8) coincides with the well-known treatment
of free-particle Brownian motion. Let us now apply numerically the
results of the present section to the Langevin dynamics of $J/\psi$ produced
in a quark-gluon plasma.

\section {\bf {Numerical Work and Discussion}}
\setcounter{equation}{0}
\renewcommand{\theequation}{4.\arabic{equation}}
{\bf 4 A. Choice of parameters}

We use $ \hbar = c =1~$units. The attractive potential 
and force between the $c \bar c$ are taken as~\cite{sve,daphne}
\begin{eqnarray}
V& =& - \frac{4\alpha_s}{3 \mid x \mid} \exp \left( - \frac{\mid x \mid}{b}
\right) \nonumber\\
f& =& \left( \frac{1}{\mid x \mid}+\frac{1}{b} \right) V~{\mbox{\rm{sign}}}(x)
\end{eqnarray}
where $\alpha_s$ is the squared QCD coupling constant, $b$ the Debye screening
length, and $b^{-1}=\mu_D$ the screening mass. Various parameters of interest have
typical values~\cite{sve} given by the following two sets :\\
\noindent Set I
\begin{eqnarray}
&&m = 0.75 ~{\mbox{\rm{GeV}}}, \quad \alpha_s =0.4\nonumber\\
&&\mu_D=b^{-1} =0.2 ~{\mbox{\rm{GeV}}}, \quad b=1~{\mbox{\rm{fm}}} \nonumber\\
&&T = 0.2~{\mbox{\rm{GeV}}}, \quad \gamma =0.081~{\mbox{\rm{fm}}}^{-1}
\end{eqnarray}
and Set II
\begin{eqnarray}
&&m = 0.75 ~{\mbox{\rm{GeV}}}, \quad \alpha_s =0.6\nonumber\\
&&\mu_D=b^{-1} =~0.2~{\mbox{\rm{GeV}}}, \quad b=~1~{\mbox{\rm{fm}}}  
\nonumber\\
&&T =0.2~{\mbox{\rm{GeV}}}, \quad \gamma =~0.157~{\mbox{\rm{fm}}}^{-1}
\end{eqnarray}
Note that $\gamma$ has been identified with the drag coefficient $A$ 
appearing in Fig.2 of Ref.~\cite{sve} and Fig.4.1 of Ref.~\cite{daphne}
after replacing an incorrect numerical factor of $1024$ by $512$.\vspace{.2in}\\

{\bf 4 B. Initial semiclassical properties} 

Strictly speaking, the variances $\Delta_{u0}^2$ and $\Delta_{x0}^2$ should
be obtained from the exact, $3$-dimensional, quantum mechanical, $s$-state
Schr\"{o}dinger wave function $\psi_0$ of the $c \bar c$ pair at $t=0$.
However, for our simple, phenomenological purpose it will suffice
to invoke the uncertainty principle for writing $|u_0| \sim \hbar/m |x_0|$.
Then the semiclassical energy reads
\begin{eqnarray}
E_0 = 1/(2m {|x_0|}^2 ) + V(|x_0|)
\end{eqnarray}
Minimization with respect to $|x_0|$ is achieved by setting $\partial E_0
/\partial |x_0| = 0$. This yields the condition
\begin{eqnarray}
1 = \frac{4m\alpha_s |x_0|}{3b} (b+|x_0|) \exp \left(-\frac{|x_0|}{b} \right)
\end{eqnarray}

which can be solved numerically to get the size $|x_0|$ in terms of $b$.
Thereby we can estimate
\begin{eqnarray}
&&\Delta_{u0}^2  \sim u_0^2 \sim \frac{1}{m^2{|x_0|}^2} \nonumber\\
&& \Delta_{x0}^2 \sim {|x_0|}^2 \quad
{\mbox{\rm {as function of}}}~b\nonumber\\
&&V_0 \sim \left( \frac{-b}{b+|x_0|} \right) \frac{1}{m{|x_0|}^2} \nonumber\\
&&\langle E_0 \rangle \sim  \frac{m}{2} u_0^2 + V_0 \sim
-\left( \frac{b-|x_0|}{b+|x_0|} \right) \frac{1}{2m{|x_0|}^2} 
\end{eqnarray}
\vspace{.2in}\\

{\bf 4 C. Time scales}

It is now straight forward to evaluate the coefficients 
$g_u$, $I_x=g_u/3$, and $g_{{}_E}=mg_u/2$ defined by Eq.(3.4). There are three
time scales {\em viz.} $\gamma^{-1}$, $\tau_x={(\Delta_{x0}^2/I_x)}^{1/3}$, and
$\tau_{{}_E} = - \langle E_0 \rangle/g_{{}_E}$ of interest in the present
problem. All our numerical results are summarized in Table 1 corresponding
to the two parameter sets (4.2 - 4.3). \vspace{.2in}\\

{\bf 4 D. Discussion}

From a physical viewpoint the time scales $\gamma^{-1}$, $\tau_x$, and 
$\tau_{{}_E}$ represent respectively the frictional relaxation, positional
swelling, and approach to ionization. A glance at Table 1 reveals that
in the case of Set I (characterized by a weaker coupling constant
$\alpha_s$ =0.4)) both $\tau_x$ and $\tau_{{}_E}$ are positive and less
than $\gamma^{-1}$. Hence Brownian movement can cause a genuine
break-up of the $c\bar c$ bound state in accordance with
Eqs.(3.5, 3.6) above.

On the other hand, in the case of Set II (characterized by a stronger
coupling constant $\alpha_s$ =0.6),  $\tau_x$ is imaginary and $\tau_{{}_E}$
are negative i.e. unphysical. Hence random force plus diffusion
cannot cause the $c \bar c$ bound state to dissociate in accordance with
Eq.(3.7) above.

Before ending we must remark that, in the context of $J/\psi$ suppression,
Langevin dynamics seems  to be almost as important as other
mechanisms (such as gluonic dissociation and Debye
screening) invoked to explain the RHIC and LHC data. Evaluation
of the charmonium survival probability~[2,6] as a function
of the transverse momentum reveals that typical time
scales corresponding to the Debye and/or gluonic mechanisms are
5 - 10 fm/c. These numbers are quite comparable to the
Langevin times $\tau_x$ and $\tau_{{}_E}$ of Table 1 (Set I) inspite
of the differences in the input parameters $T$ and $\mu_D$.

\section {\bf {Additional Complications}}
\setcounter{equation}{0}
\renewcommand{\theequation}{5.\arabic{equation}}

We now examine critically the validity of several 
oversimplifications done above and also point out some additional
complications likely to arise in future applications of the theory :

(a) {\underline{Bound state in 3-dimensions}}:

One may argue that the simple, 1-dimensional, uncertainty principle
based treatment of Eqs. (4.4, 4.5) will break down for the real charmonium
which is a bound state in 3-dimensions. To answer this, we replace $|x_0|$
by $r_0$ (which is the absolute distance between the $ c \bar c$ pair)
and appeal to the semiclassical, circular, Bohr orbits picture
analogous to the familiar hydrogen atom problem. The ground
state orbit in a Yukawa force $f$ defined by Eq. (4.1) has principal
quantum number $n=1$, orbital angular momentum $L_0 = \hbar =1$,
and centrifugal force condition
\begin{eqnarray}
\frac{m u_0^2}{r_0} = \frac{L_0^2}{m r_0^3} = - f_0 
\end{eqnarray}
which is entirely equivalent to the transcendental equation (4.5).
It follows that our earlier estimate (4.6) of the bound state energy remains
correct even for the real charmonium.

(b) {\underline{3-dimensional random walk}}:

 Next, it is worth asking
whether the main results of Secs. 2, 3 will get drastically altered
if the random walk occurs in actual 3 dimensions. To answer this,
we look at the vector Langevin equation
\begin{eqnarray}
\frac{d \vec{u}}{dt} + \gamma \vec{u} = (\vec{f} +\vec{h})/m
\end{eqnarray}
where formal solutions, Taylor expansions, and stochastic averaging
may be done on the same lines as Eqs. (2.6 - 3.2). Modifications
appropriate to 3 dimensional configuration space can be readily done
at every step. For example, at $t=0$, we would have $\langle \vec{f_0} 
\rangle = \vec{0}$, by parity argument while $ \langle \vec{r_0}. \vec{f_0}
\rangle /m = - \langle {\vec{u_0}}^2 \rangle$ by the virial theorem. The
crucial point to be noted is that the velocity variance $\Delta_{\vec{u}}^2$
will increase with $t$ linearly and the position variance $\Delta_{\vec{r}}^2
$ will do so cubically at small times, i.e., the essence of our leading 
behaviour (3.3) would remain intact.

(c) {\underline{Choice of initial conditions}} :

One may claim that the time $t=0$ should be set at the instant when the
$c \bar c$ pair was created in the plasma by a hard partonic collision
having divergent trajectories. In other words, the Langevin dynamics ought
to have been applied even to the ``pre-resonance formation stage'' where 
some energetic pairs would lose their excess energy by random walk to form
a bound cluster, not necessarily an $s$- wave ground state. This view, though
very correct and ambitious, has three practical difficulties. First,
pioneers like Xu et al and Karsch [2] have not adopted this view.
Second, the initial values of 
\begin{eqnarray}
\langle x_0 \rangle, \langle u_0 \rangle , \langle x_0 u_0 \rangle
\end{eqnarray}
are not known immediately after the hard partonic reaction. Third,
the final Brownian variances based on Eq. (5.3) will contain a large
number of undetermined coefficients.

The initial conditions at $t=0$ imposed in the present work correspond
to a ``{\em fully - formed $c \bar c$ discrete bound state}''. This view, though
modest, has three practical advantages. First, some pioneers of gluonic
dissociation [2] have taken the initial state to be a standard $c \bar c$
resonance like $\psi$, $\psi^\prime$, $\chi$ etc. Second, the stochastic
inputs $\langle x_0 \rangle = \langle u_0 \rangle = \langle x_0 u_0 \rangle
= 0 $ are precisely known. Third, the final variances in Eq. (3.3) involve 
only one known effective coefficient $g_u$.

(d) {\underline {Use of barycentric frame}} :

 One may raise the criticism that there is no freedom to go to the 
$c \bar c$ barycentric system because the plasma - which is the source
of random noise - provides a fixed frame of reference. For a plasma 
{\em at overall rest} this criticism is readily met by remembering
that the noise $h(t)$ being a function of time is Galilean invariant.
Indeed, in a general frame of reference, the pair Hamiltonian
$H_{12}$ reads
\begin{eqnarray}
H_{12} = p_1^2/2m_1 + p_2^2/2 m_2 + U_{QGP} + V(x_1 - x_2) - x_1 h_1 - x_2 h_2
\end{eqnarray}
where $U_{QGP}$ is a constant potential generated in a spatially
homogeneous plasma. Obviously, a separation between the centre of mass
coordinate $(x_1+x_2)/2$ and relative coordinate $(x_1 - x_2)/2$ can be
effected in Eq. (5.4).

There is, however, an important word of caution here. In reality, the plasma
{\em evolves rapidly with the time} by virtue of longitudinal/transverse
expansion. The above-mentioned passage to the $c \bar c$ centre
of mass frame is justified in Bjorken's boost-invariant hydrodynamics [6]
if the $c \bar c$ pair moves either longitudinally or has small
transverse momentum $p_T$. The procedure, however, may not be 
justifiable for large $p_T$ pairs. This is because the pair distribution
should relax to the equilibrium form in the plasma rest frame
which would look different in other frames.

(e) {\underline {Miscellaneous refinements}} :

There are a few other subtle points to which attention will have to be
paid in future. Since the $c \bar c$ pair may be in a colour singlet
or octet state [7] a coupling between these channels may occur in the
equation of motion. Next, the possibility of having different diffusion
constants along the longitudinal and transverse directions should
be allowed so that the asymptotic equilibrium distribution of the
charmed quark acquires a Tsallis shape [11] instead of the Boltzmann
form. Finally, since the initial state of the charmonium is
necessarily quantum-mechanical, a path-integral based density
matrix may be formulated by taking hints from single-particle [12]
or multiparticle [13] quantum stochastic dynamics.

\section*{Acknowledgements} We thank Dr. Dinesh Srivastava for 
useful discussions during the early phases of this
work.

\newpage
\medskip
%%%%%%%%%%%% begin tab,1%%%%%%%%%%%%%%%%%%%%%%%%%%%%%%%%%%%%%%%
\begin{table}[h]
\caption{ Numerical results on $J/\psi$ suppression
due to Brownian movement in the barycentric frame. For
notations see the text.}
\vskip 0.2in
%\centerline{
\begin{tabular}{|l|l|l|l|l|l|l|l|}
\hline
  & & & & & & &  \\
 Set & $|x_0|$ & $\Delta_{u0}^2$ & $\Delta_{x0}^2$ &$\langle E_0 \rangle$
 &$\gamma^{-1}$ & $\tau_x$ & $\tau_{{}_E}$ \\
& (fm) & $(c^2)$ & $(fm^2)$ & (GeV) & (fm/c) & (fm/c) & (fm/c) \\
%  && & & & &  & \\
\hline
& && & & & & \\
$I$  & 0.553& 0.226 &0.306  &-0.0245  &12.279 &5.175 &9.8447\\
 && & & &  & &\\
\hline
& && & & &  &\\
$II$ &0.346 &0.579 &0.119 &-0.107 &6.37 & ${(-3.64)}^{1/3}$ &-2.91\\
\hline
\end{tabular}
%}
\end{table}

\end{document}